\documentstyle[12pt]{article}

\input epsf

\begin{document}

\renewcommand{\thesection}{\arabic{section}.} 
\renewcommand{\theequation}{\thesection \arabic{equation}}
\newcommand{\scs}{\setcounter{equation}{0} \setcounter{section}}
\def\req#1{(\ref{#1})}
\newcommand{\be}{\begin{equation}} \newcommand{\ee}{\end{equation}} 
\newcommand{\ba}{\begin{eqnarray}} \newcommand{\ea}{\end{eqnarray}} 
\newcommand{\la}{\label} \newcommand{\nb}{\normalsize\bf} 
\newcommand{\lb}{\large\bf} \newcommand{\vol}{\hbox{Vol}}
\newcommand{\bb} {\bibitem} \newcommand{\np} {{\it Nucl. Phys. }} 
\newcommand{\pl} {{\it Phys. Lett. }} 
\newcommand{\pr} {{\it Phys. Rev. }} \newcommand{\mpl} {{\it Mod. Phys. Lett. }}
\newcommand{\sg}{{\sqrt g}} \newcommand{\sqhat}{{\sqrt{\hat g}}}
\newcommand{\sqphi}{{\sqrt{\hat g}} e^\phi} 
\newcommand{\sqalpha}{{\sqrt{\hat g}}e^{\alpha\phi}}
\newcommand{\tp}{\cos px\ e^{(p-{\sqrt2})\phi}} \newcommand{\stwo}{{\sqrt2}}
\newcommand{\tr}{\hbox{tr}}

\begin{titlepage}
\renewcommand{\thefootnote}{\fnsymbol{footnote}}

\hfill CERN--TH/2000--316

\hfill hep-th/0011065

\vspace{.4truein}
\begin{center}
 {\LARGE Strings from Logic}
 \end{center}
\vspace{.7truein}

 \begin{center}

 Christof Schmidhuber\footnote{christof.schmidhuber@cern.ch}

 \vskip12mm

 {\it CERN, Theory Division, 1211 Gen\`eve 23, Switzerland}

 \end{center}

\vspace{1.5truein}
\begin{abstract}\vskip5mm
\noindent
What are strings made of?
The possibility is discussed that strings
are purely mathematical objects,
made of logical axioms. More precisely,
proofs in simple logical calculi are 
represented by graphs that can be interpreted as the Feynman diagrams of
certain large--$N$ field theories. Each vertex represents an axiom.
Strings arise, because
these large--$N$ theories are dual to string theories.
These ``logical quantum field theories'' map theorems into the space of functions
of two parameters: $N$ and the coupling constant. 
Undecidable theorems might be related to nonperturbative
field theory effects. 

\vskip1cm\noindent
{\it Based on a talk given at CERN (Nov 7, 2000)}
\end{abstract}
 \renewcommand{\thefootnote}{\arabic{footnote}}
 \setcounter{footnote}{0}
\end{titlepage}

\subsection*{1. Introduction}

\vskip5cm

\epsffile[-5 5 0 0]{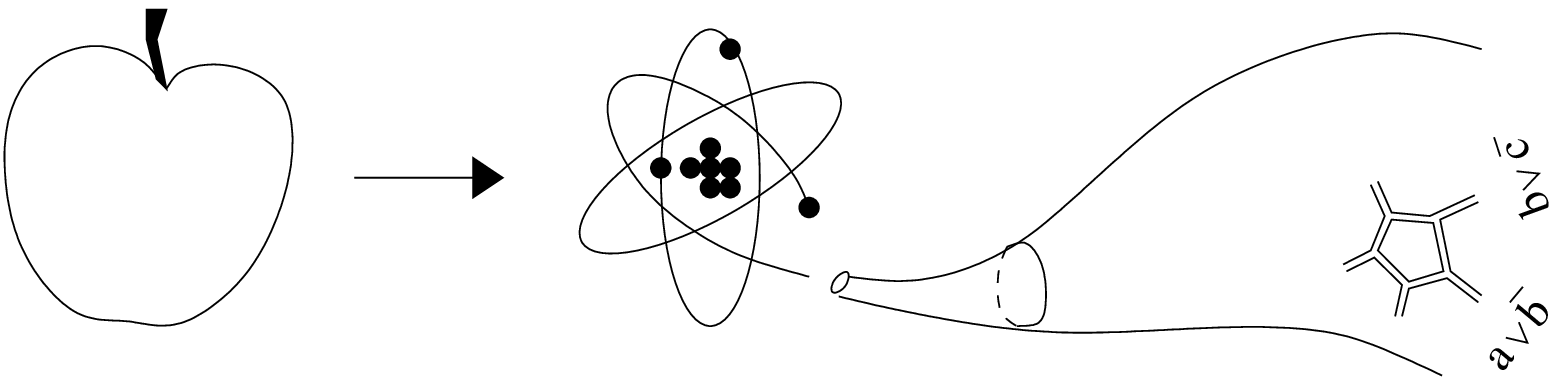}

\begin{center}
\end{center}

\noindent
Apples are made of atoms, atoms are made of elementary particles, and
elementary particles are presumably excitation modes of superstrings.
It is natural to ask what superstrings are made of.
More generally, it is natural to wonder whether this chain of questions 
for more and more fundamental constituents of 
 the things we observe 
can ever end, or whether it
goes on ad infinitum.

Here I would like to present some observations that can perhaps
be regarded as a realization of
an ancient and vague idea how
the chain of questions might end. 
According to 
 this ancient idea, 
 the copuscles of the four elements 
(water, air, earth and fire) represent objects of
pure mathematics (such as regular polyhedra) \cite{pla}. The 
 chain of questions ends, because
it is meaningless to further ask for the origin of such mathematical objects
-- they ``are simply there''.
The present proposal does not involve polyhedra.
But it involves
string world--sheets that are made
of axioms of mathematical logic in a sense that will be made precise. 

These ``logical'' string world--sheets
are what one stumbles upon in the course of pursuing a
different project that at first seems 
unrelated to string theory. Its 
starting point is the standard
formalization of branches of mathematics such as number theory:
one introduces variables that represent propositions $p\in\{true, false\}$,
variables that represent numbers $n\in\{1,2,3,...\}$, and symbols 
such as $\wedge,\vee,\neg,\exists,=,+,...$ out of which new propositions
 are formed.
Starting from a basic set of axioms (propositions that are
defined to be true), theorems are constructed (``proven'')
by manipulating the symbols according to given ``rules of inference''.
By applying the axioms in all possible ways,
one obtains the set of all theorems that can be proven within this
formal system.

The project mentioned is to look for
what in the following will be called ``logical quantum field theories'':
field theories, whose 
Feynman diagrams graphically
represent the proofs in a given formal system, with each vertex  
representing a single axiom.
We'll find that at 
least for some very
simple subsystems of the propositional and predicate calculus
such quantum field theories exist. They are well--known
zero--dimensional $SU(N)$ field theories whose coupling constant will be called
$\beta$.
Their correlation functions correspond to
theorems and are
\begin{enumerate}
\item
 zero, if the theorem is false;
\item
 $e^{-\beta L}\ p(N)$ \ for $\beta\rightarrow\infty$ \ if the theorem is true,
\end{enumerate}
 where $L$ is the ``length'' of the shortest proof and $p(N)$ is a 
power series.
These theories thus provide a map from
theorems to functions of $\beta$ and $N$. 
One original motivation for looking for such field theories was
to relate undecidable theorems that are true but can not be proven within
a given formal system to 
features of field theory that are nonperturbative in $\beta$ 
(and perhaps in $N$). 
This will only be mentioned here and will be worked out elsewhere.

My emphasis here is on
 a different aspect of  these logical quantum field theories:
they are dual to string theories;  the 
string world--sheets are built out of the axioms of the
formal system. 
I will also  indicate how more complicated formal 
systems might involve {\it super}strings
and membranes. 
This will then raise the question whether these logical string world--sheets
can be identified with 
the physical ones, thereby  motivating the picture drawn above.

In section 2, the basic idea is explained at a primitive
toy model of a formal system. Section 3 begins
to generalize this model to the
standard propositional calculus. After it is explained
what the propositional calculus has to do with strings, open
issues are discussed in section 4. It is
suggested how membranes and superstrings
 may arise, and
unprovable theorems are also commented on. 
Section 5 then tries to take seriously the proposal that
the {\it physical} strings could be such logical strings.

The ideas that I will mention here
 are part of more general views that have developed
 over the past 20 years. They were particularly stimulated
by discussions with J\"urgen Schmidhuber, whose study of the ensemble
of computable bit--sequences (interpreted as desribing universes)
is inspired by similar thoughts \cite{jur}.
As I noticed recently, Tegmark \cite{teg} also advocates
identifying the physical world with a 
purely mathematical one, and suggests that logical calculi
play some role (see also earlier books and articles cited in \cite{teg}
that point in this direction).
Our observations may be regarded as a 
concrete (and stringy) realization of such general suggestions.
\newpage

{}\subsection*{2. Toy Model}\scs{2}

First, I would like to explain the basic ideas
at the example of a primitive toy model. 
Theoreticians who are familiar with
the ``old matrix model'' will find that I essentially
reinterpret it in terms of logic. 
The reason for doing this will become clear in section 3:
there, the toy model will serve
as a prototype for part of the standard propositional calculus.
In order to get to the point quickly, the style of this section
is a bit  sloppy; definitions of symbols and phrases
 are omitted if their use is
intuitively clear.

\noindent
Our toy formal system
consists of the following elements:
\begin{enumerate}
\item
Variables $a,b, ...$ which we assume to be already defined (denoting, e.g., numbers).
\item
A predicate symbol ``=''.
\item
Two axioms:\ \ If ($a=b$ and $b=c$) , then $a=c$ 
$$\hskip16mm
\hbox{If}\ a=c\ ,\ \hbox{then there is a value of \ $b$ \ such that}\ 
       (a=b\ \hbox{and}\ b=c)\ .$$
\end{enumerate}
We want to
\begin{itemize}
\item[(i)]
identify a ``logical 
quantum field theory'' whose Feynman rules
represent the axioms, whose Feynman diagrams represent proofs,
and whose correlators represent theorems.
\item[(ii)]
illustrate in what sense the axioms are the constituents of string
world--sheets.
\end{itemize}
 To this end, let us graphically represent the axioms in terms of double line
diagrams
as shown in the figure (left and center). Each single line represents a variable
and each pair of lines represents a proposition such as $a=b$.
The diagrams are read from top to bottom. In this sense, the two
axioms are just ``time-reversed'' interpretations of the same graph.

\vskip4cm
 
\epsffile[-5 5 0 0]{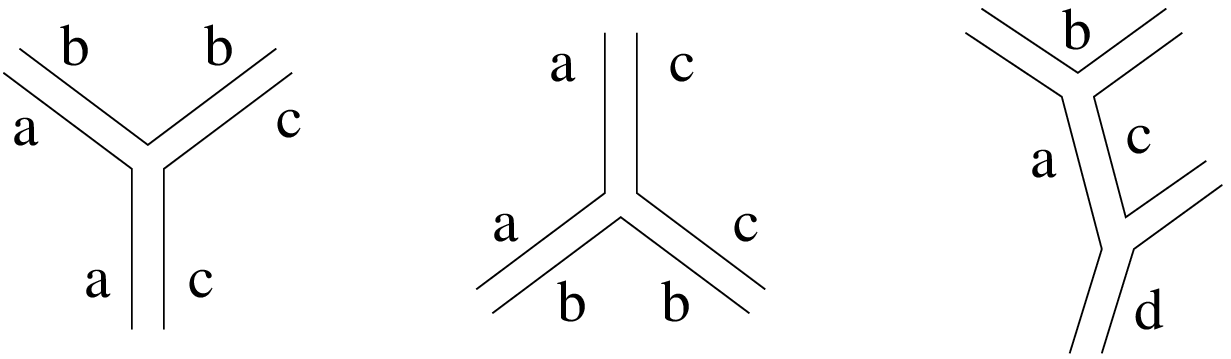}

\vskip1cm
\begin{center} 
{\small Figure 2: Two axioms and a proof.}
\end{center}

\noindent
By gluing axioms together, ``theorems'' can be proven.
The right graph shows an example of a proof that uses
the first axiom twice:
\ba
\hbox{Axiom:}&&
\hbox{If}\ \ (a=b\ \hbox{and}\ b=c)\ ,\ \ \hbox{then}\ \ a=c\\
+\ \hbox{Axiom:}&&\hbox{If}\ \ (a=c\ \hbox{and}\ c=d)\ ,\ \ \hbox{then}\ \ a=d\\
\rightarrow\ \hbox{Theorem:}&& 
\hbox{If}\ \ (a=b\ \hbox{and}\ b=c\ \hbox{and}\ c=d)\ ,\ \ \hbox{then}\ \ a=d
\la{aachen}\ea
Note that there are different theorems that are represented by the same graph.
E.g., the right graph can also be read ``crossed'', such that it represents
 a proof of the theorem
$$ \hbox{If}\ \ (b=a\ \hbox{and}\ a=d)\ ,\ \ \hbox{then} \ \ 
\hbox{there is a}\ \ c\ \ \hbox{such that}\ (b=c\ \hbox{and}\ c=d)\ .
$$
We now imagine a ``proof machine'' 
 that systematically derives all true statements
by performing all possible successive applications of the axioms.
This machine will stupidly produce many redundant proofs, such as
\ba
\hbox{Axiom:}&&\hbox{If}\ \ a=b\ ,\ \ \hbox{then}\ \ \hbox{there is a}\ \ c\ \ \hbox{such that}\ \ 
(a=c\ \hbox{and}\ b=c)\\
+\ \hbox{Axiom:}&&\hbox{If}\ \ (a=c\ \hbox{and}\ b=c)\ ,\ \ \hbox{then}\ \ a=b\\
 \rightarrow\ \hbox{Theorem:}&& \hbox{If}\ \ a=b\ ,\ \ \hbox{then}\ \ a=b\ .
\la{dublin}\ea
This proof corresponds to a loop diagram (see figure 3, left)
with an auxiliary variable $c$ inside the loop.
By gluing the axioms together in all possible ways
one obtains the set of all theorems and proofs, 
represented in terms of double line graphs.
One of them is shown in figure 3 (center).

\vskip5.5cm
 
\epsffile[-5 5 0 0]{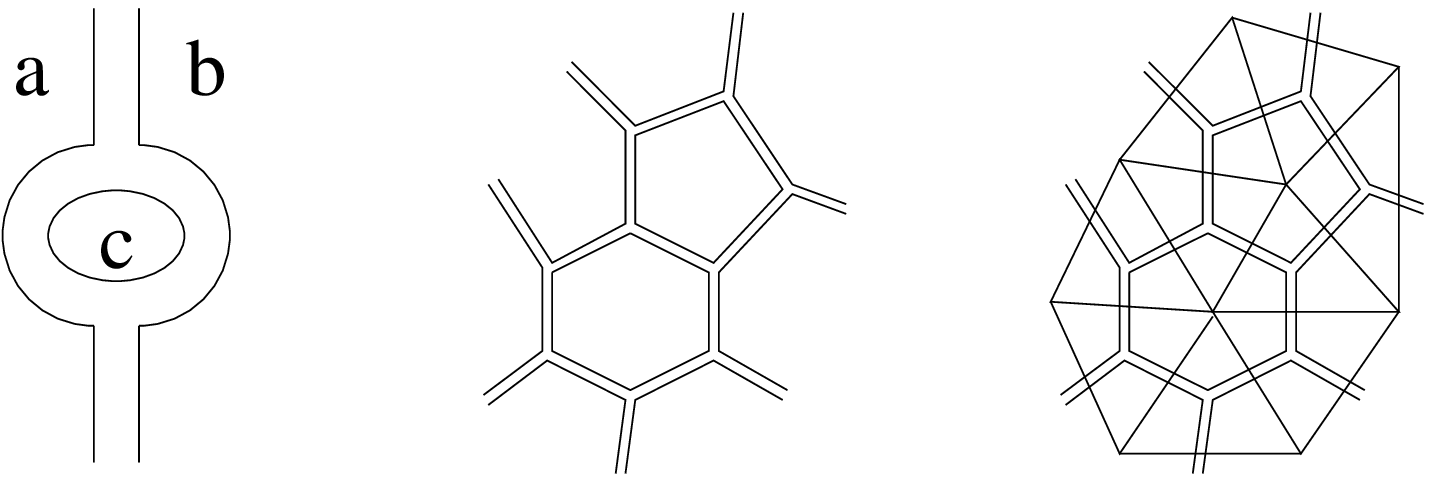}

\vskip1cm
\begin{center} {\small Figure 3: More proofs.}
\end{center}

All theorems that can be proven by our machine are of course completely
trivial. This should not bother us, because in this note we 
are not  interested in any nontrivial content
of individual theorems. Instead, we are going to consider 
the set of theorems and proofs in simple formal systems as
statistical--mechanical ensembles and study
properties of the ensembles. This is a rough analog of the situation in
thermodynamics, where one does not care about the shape of the individual
gas molecules. One only cares about statistical properties of the
ensemble of gas molecules such as 
temperature, pressure, etc.
In the case at hand, the
ensemble of proofs has two
 equivalent representations, which we now discuss. 

First, our double--line diagrams are well--known to be the dual graphs
of random triangulations of a
 two--dimensional surface with boundary \cite{tho}
-- i.e., of a Euclidean open string world--sheet (figure 3, right). 
The vertices 
of the double--line graphs
correspond to the centers of the
triangles, and the external double lines (which make up the theorem
to be proven) correspond to the boundary 
of the triangulation.
The diagrams drawn so 
far are ``planar'' in the sense that the dual surface has the topology
of a disk.
It is possible to interpret higher--genus diagrams in terms of
proofs in which some of the external or internal variables are identified.
Note that  
theorems in which an external variable appears an odd number
of times cannot be proven within this formal system.

Second, our
double line graphs
are of course also well--known to be the Feynman diagrams of a
zero--dimensional field theory of a real symmetric
 $N\times N$ matrix $M$ with Lagrangean \cite{itz}
\ba{\cal L}\ =\ N\ \tr (M^2\ +\ e^{-\beta} M^3)\ .\la{mxm}\ea
Here, the ``coupling constant''
$\beta$ is a free real parameter and the number $N$ of
SO($N$) ``colors'' is a free integer parameter.
This is the ``old matrix model'', which is the
 ``logical quantum field theory'' for our toy formal system.
The value of a Feynman diagram with $L$ vertices,
representing a proof with $L$ axiomatic steps, is well--known to be
$e^{-\beta L}N^{2-2g}$, where $g$ is the genus of the corresponding surface.
So the
correlation function (or Wilson loop)
 corresponding to a theorem $T$ has an expansion of 
the form
\ba\ f_T(\beta,N)\ \sim\ \sum_{p(T)}\ e^{-\beta L(p)}\ N^{2-2g(p)},\la{sum}\ea
where $p(T)$ runs over all inequivalent proofs of $T$.
In the ``classical limit'' $\beta\rightarrow\infty$, 
the shortest proof dominates
as advertised in the introduction.

Alternatively, $\beta$ can be
 fine--tuned  to a certain critical
value $\beta_0$ (as is also well--known), 
where the continuum limit is reached
and large proofs just begin to dominate in (\ref{sum}).
In this limit,
the ensemble of the random triangulations that are dual to our proofs
is in the same universality class
as the ensemble of continuum string world--sheets \cite{kaz,gro}.
(\ref{sum}) becomes a correlation function
in a non--critical Euclidean string theory, whose
 string world--sheets are literally ``made of axioms''. 
In section 5, it will be discussed whether these
 logical strings can be identified with the physical strings. 

To summarize: in the case of the toy model,
the logical quantum field theory 
is a zero--dimensional SO($N$) theory
with coupling constant $\beta$.
It defines a map from theorems into the space of functions $f(\beta,N)$ of
two parameters. These functions are
the corresponding correlation functions. The situation is,
at least superficially, analogous to the situation in topological
quantum field theories, where knots and links are mapped onto
 functions of two parameters $k$ and $N$ \cite{wit} (which motivates
the name ``logical quantum field theories'').

{}\subsection*{3. Propositional calculus}\scs{3}
 
In this section we begin to
 generalize the discussion to the standard propositional
calculus, which is a basis for
 most of the interesting
higher calculi, such as number theory.
The predicate calculus is commented on in the next section.

\subsubsection*{3.1. Setup\footnote{All the background that we need
here is contained in the brief review in the
Encyclopedia Britannica.}}

 The  propositional calculus
contains only propositions and no numbers or
other variables. 
Propositions will be denoted by $a,b,c,...\in\{true,false\}$.
New propositions such as 
$$\neg\ a\ ,\ \ \ \ \ a\ \ \vee\ \ b$$
can be built from old ones via $\neg$ (``not''),
 $\wedge$ (``and''), $\supset$ (``implies'')
and $\vee$ (``or''), which we take to mean ``either $a$ or $b$ or both''.
We denote the negation of $a$ by $\bar a\equiv \neg a.$
Brackets (,) and quantifiers $\exists,\forall$
will be used in the standard way.
$\wedge$ and $\supset$ may be eliminated as independent symbols
by expressing them
in terms of $\vee$ and $\neg$:
$$a \wedge b\ \equiv\ \neg (\bar a\vee \bar b)\ \ \ ,\ \ \ 
    a \supset b\ \equiv\ (\bar a\vee b)\ .$$

In the propositional calculus, statements can always be
decided via ``truth tables'': a statement is a theorem if it is true
for all values (true or false) of all the variables it consists of,
e.g. $\bar a\vee a$.
In more complicated calculi, where variables can, e.g., be numbers,
this is generally not possible,
since it would take an infinite amount of time
to check infinitely many possible values of the variables.
In this case one needs to {\it prove} theorems using
axioms and rules of inference. This approach can of course
be used even for the propositional calculus.

For the propositional calculus,
there are many equivalent formal systems, i.e., sets of axioms and
rules of inference that are {\it complete}
(all true statements of the propositional calculus can be derived) 
and {\it sound} (no false statements can be derived).
The system that will be used here is a variation of
that of Russell and Whitehead \cite{rus}.
The latter contains 4 axioms\footnote{The original system contained
5 axioms, but the fifth one was later derived from the other 4.}.
The first three are:
\ba
\hbox{Axiom 1:}\hskip15mm&& (a\ \vee\ a)\ \ \supset\ \ a\\
\hbox{Axiom 2:}\hskip15mm&& a\ \ \supset\ \ (a\ \vee\ b)\\
\hbox{Axiom 3:}\hskip15mm&& (a\ \vee\ b)\ \ \supset\ \ (b\ \vee\ a)\ .\la{A3}\ea
The fourth and most interesting axiom is
$$(b\ \supset\ a)\ \ \supset\ \ [(\bar c\ \vee\ b)\ \supset\ 
(\bar c\ \vee\ a)]\ .$$
To make the relation with strings more manifest (see below), it is useful to
rewrite it as
\ba\hbox{Axiom 4:}\hskip20mm \{(a\vee\bar b)\ \wedge\ (b\vee\bar c)\}
\ \supset\ (a\vee\bar c)\ .\la{mick}\ea
In rewriting this axiom, axiom 3 and
associativity of $\vee$ must be used. Associativity
can be proven with the original 4 axioms,
but we have not succeeded in proving it with the new 4th
axiom instead of the old one. To be safe, we thus add associativity as a new
axiom:
\ba\hbox{Axiom 5:}\hskip15mm&& \{(a\vee b)\ \vee\ c\} \ \ \supset\ \ 
\{a\ \vee\ ( b \vee c)\} \ .\la{asso}\ea
In addition to the axioms, there are two rules of inference:
\begin{enumerate}
\item
the {\it modus ponens:}
if $A$ and $A \supset B$ are theorems, then $B$ is a theorem.
\item
the rule of 
 {\it substitution:}
 if $P(a)$ is a theorem, then $P(f)$ is a theorem, where
$a$ is replaced
by any composite (but not necessarily true) proposition $f(a_1, a_2, ...)$.
\end{enumerate}
Similarly as in section 2, the goal is to 
construct ``logical quantum field theories''
whose Feynman diagrams represent the proofs
that can be constructed from axioms 1--5 using these two rules.
We will proceed as follows.
In subsection 3.2, proofs will be discussed
that can be built from only the fourth axiom using the modus ponens.
In subsection 3.3, axioms 1--3
will be incorporated. Section 4 contains open issues:
subsection 4.1 speculates how to
include the rule of substitution and axiom 5, which are crucial
for proving interesting theorems.
Subsection 4.2 adds elements of the lower predicate calculus. Subsection 4.3
suggests a relation between undecidable theorems 
and nonperturbative field theory.

{}\subsubsection*{3.2. Axiom 4}

To see how strings arise in the propositional calculus, we first
consider only the subset of theorems that
can be proven using axiom 4 (\ref{mick})  alone
(and its ``time reversal'' -- see below).
Let us graphically represent this axiom 
in terms of double lines
as shown in figure 4 (left).

\vskip5cm
 
\epsffile[-5 5 0 0]{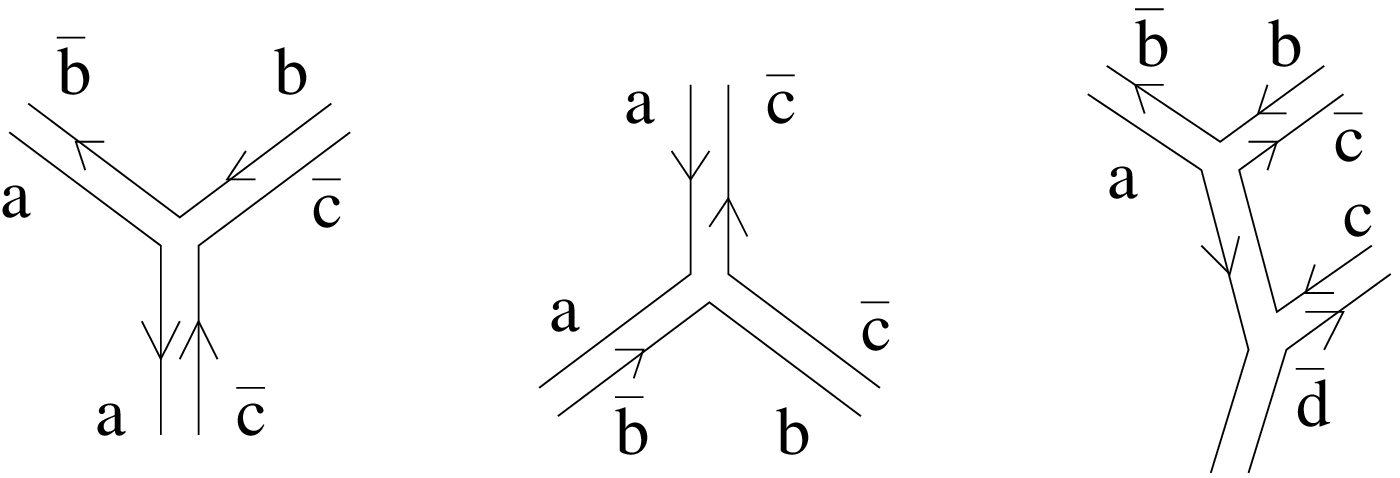}

\vskip7mm
\begin{center} {\small Figure 4: Axiom 4.}
\end{center}

\noindent
The difference
to the diagram for the toy model in figure 2 is that each line now represents a
proposition $a,b,...\in\{true,false\}$ rather than a general
variable such as a number.
Also, the lines carry arrows. Let us define ``time'' as the direction
in which the diagram is read.
Arrows that run backwards in time
are understood to represent the negated propositions.
 If time runs from top to bottom, axiom 4 is recovered.
The diagram still represents a true statement, though,
if time runs in any other direction:
reading the graph from the lower left to the upper right yields 
 the same axiom with
$a,b$ and $c$ permuted; reading the graph from bottom to top
yields another true statement, provided that we insert an ``$\exists$'': 
$$(a\vee\bar c)\ \supset\ \exists b:\ (a\vee\bar b)\ \wedge\ (b\vee\bar c)\ .$$
This is drawn in figure 4 (center).
(The introduction of the symbol $\exists$, which is usually reserved for the
predicate calculus, is convenient but
 not really necessary here: in the propositional
calculus, $\exists b:f(b)$ can be written out as $f(1)\vee f(0)$.)

The modus ponens implies that, as
in the toy model, new theorems such as
$$ \ \{(a\vee\bar b)\ \wedge\ (b\vee\bar c)\ \wedge\ (c\vee\bar d)\}
\ \ \supset\ \ (a\vee\bar d)\ $$
can now be proven by 
gluing together axioms as shown in figure 4 (right). 
If there are closed lines (similarly as in figure 3, left), 
the corresponding variable $c$
in the loop is assumed to come with an $\exists$.
Again, by combining graphs representing axiom 4
in all possible ways, one can derive all theorems that can be
proven with this axiom alone. 
This yields a small and rather trivial subset of all
theorems of the propositional calculus. 
As in the toy model, the corresponding diagrams are the dual graphs
of random triangulations of a two--dimensional surface with boundary
(as in figure 3).
The only difference is that each line now has two possible orientations.

The logical quantum field theory whose Feynman diagrams
are the proofs built from axiom 4 is again an ``old matrix model''
with some coupling constant $\beta$,
but the two possible orientations translate into having
complex instead of real symmetric matrices.
The color indices
$a,b,c\in\{1,2,...,N\}$ that label the single lines
are now $SU(N)$ indices.
Correlation functions again correspond to theorems
and are functions of $\beta$ and $N$ of the form (\ref{sum}).

{}\subsubsection*{3.3. The first three axioms}

So far, only the fourth axiom has been graphically represented.
How about the first three axioms of Russell and Whitehead?
As for axiom 3 (\ref{A3}), it is already implicit in our formalism,
since the double lines can be twisted (figure 5, left).
For our random surfaces, this possibility of twisting implies
that they are in general not orientable.
As for axioms 1 and 2, those can be written as
\ba
\hbox{Axiom 1:}&&(a\ \vee\ a)\ \ \supset\ \ (a\vee 0)\ \wedge\ (1\vee a)
\la{a1}\\
\hbox{Axiom 2:}&&(a\vee0)\ \wedge\ (1\vee b)\ \ \supset\ \ (a\ \vee\ b)\la{a2}
\ea
Here we use
$a\wedge1=a,a\vee0=a,a\vee1=1,a\wedge0=0$. We
regard these four statements as auxiliary axioms,
 defining ``0'' and ``1''.
 In the form (\ref{a1}),(\ref{a2}),
axioms 1 and 2 become analogous to axiom 4.
This is also drawn in figure 5 (second and third graph), where the dashed line
represents ``0'' or its negation ``1''.

So it seems that it is not difficult to include
the first three axioms in our proof diagrams: this involves
dashing some of the internal or external lines,
so that they denote 0's or 1's. 
Note that with the help of 0's and 1's in external lines
we can now also prove theorems in which a
given variable occurs an odd number of times.
Note also a curious aspect of
axiom 2: it
 violates ``time reversal'' in the sense that it cannot be read from
bottom to top: there
is no theorem that says $(\bar a\vee\bar b)\supset (\bar a\vee1)\wedge
(0\vee\bar b)$,
except when $a=b$ as in the first axiom.
It remains to be understood what this means for the string 
world--sheets.
Note finally that the modus ponens, if applied to single lines
rather than double lines, 
is also contained in our diagrams (figure 5, right), since it
 can be written as
$$(0\vee A)\ \wedge\ (\bar A\vee B)\ \supset\ (0\vee B)\ .$$

\vskip4cm
 
\epsffile[-5 5 0 0]{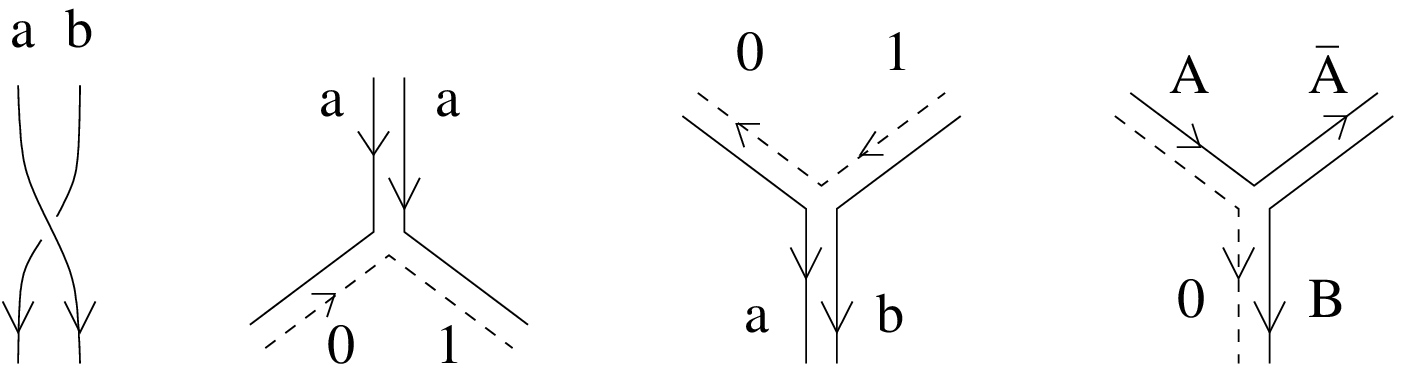}

\vskip10mm
\begin{center} {\small Figure 5: Axioms 3, 1, 2, and the modus ponens.}
\end{center}

{}\subsection*{4. Open issues and conjectures}\scs{4}

In this section, three open problems and their possible solutions
are discussed:
\begin{enumerate}
\item
It remains to graphically represent
axiom 5 (\ref{asso}) and the rule of substitution.
As I will explain, this might
require extending the strings to membranes.
\item
The discussion should be generalized to the lower predicate
calculus. I will suggest that this involves higher--dimensional
string embedding spaces and may involve superstrings.
\item
How do undecidable theorems show up in our formalism? 
I will speculate that they 
show up as nonperturbative effects in the logical quantum field theories.
\end{enumerate}

{}\subsubsection*{4.1. Axiom 5 and the rule of substitution}

So far, our proof diagrams  do not account for the 
the possibility
that a proposition variable that is part of a theorem
may be substituted by a composite proposition. 
Without including this ``rule of substitution'', we can only prove a tiny
and rather trivial
subset of theorems of the propositional calculus.
In terms of the double--line diagrams of Fig. 4,
substitution means that  
the propositions $p$ and $q$
 in a double line that represents $p\vee q$
 may be composed of more elementary propositions.
In other words, a closer look at any of the single lines may reveal
that the single line is itself a pair of lines, or a pair of
pairs of lines, and so on. 

A natural way to represent this graphically
is the following.
We first use the fact that every
well-formed (but not necessarily true) proposition
 can be drawn as an electric cirquit. The example $p\vee q$ with
$p=(a\vee b)$ and $q=\{b\vee\neg(c\vee d)\}$ is drawn in figure 6, left. 
In the short--hand notation of figure 6 (center), the cirquit becomes a 
branched tree: the 
negation is represented by a dashed line and
the 3--vertex represents either $\vee$
or a branching in the inputs $a,b$, depending on whether it branches
upwards or downwards.

\vskip5cm
 
\epsffile[-5 5 0 0]{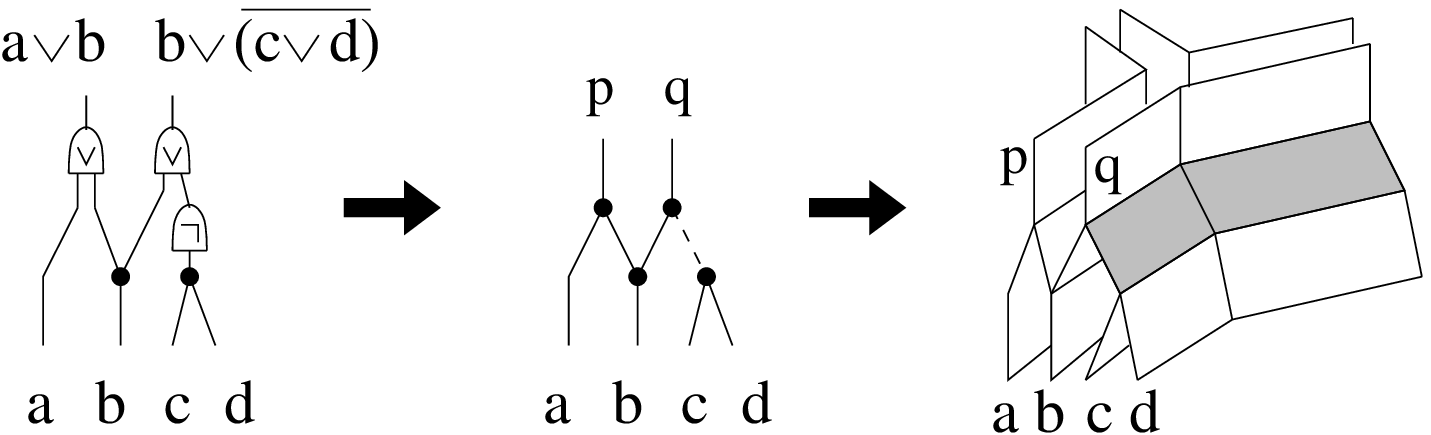}

\vskip10mm
\begin{center} {\small Figure 6: Substitution.}
\end{center}

Let us now introduce a third dimension in the planar Feynman diagrams
of Fig. 4,
perpendicular to the plane on which the planar diagram is drawn 
(figure 6, right). We extend the double lines of the planar diagram
into the third dimension. This is done such
that they become branched sheets,
a cross section through which is the branched tree just
constructed (dashed lines are represented in the figure by shaded areas). 
In this representation, the sum over proofs becomes
a sum over three--dimensional branched structures.
But these structures can not only
branch in the direction of the third dimension. They can also branch in
the directions parallel to the original plane,
in the sense that the trees can be manipulated 
during the proof process, e.g., by axiom 5:
$$ (p_1\vee p_2)\ \vee\ p_3
\ \ \ \rightarrow\ \ \  p_1\ \vee\ (p_2 \vee p_3)\ .$$
The open problem is now to describe the statistical--mechanical
properties of the ensemble of such branched sheets. This looks like a
tough problem, but there is an immediate conjecture: that the ensemble
of such three-dimensional graphs is effectively represented by dynamical
membranes; and that the associated ``logical quantum field theory''
is related to M--theory.\footnote{A perhaps related  hint at 
membranes is the observation
 that general propositions can be
written as $(a_1\vee a_2\vee a_3)\wedge(b_1\vee b_2\vee b_3)\wedge...$
and can be represented by triple--line (rather than double--line) graphs.
Those may be dual to discretized membranes.}
Of course, even if this can be made precise, one is left with the
usual problems of ``random membranes'': the sum over three--dimensional 
topologies is not understood\footnote{Although a subclass of topologies,
Seifert manifolds, can be described in a field theory context \cite{cstop}.},
and there is no renormalizable theory of dynamical three--dimensional
gravity, so we cannot expect anything like a second--order phase transition.
We must leave the subject for the future; perhaps
looking for
relations with matrix theory \cite{mx} may help.

{}\subsubsection*{4.2. Lower predicate calculus}

In the propositional calculus, a proposition can be built out of other
propositions, such as $a\vee b$. 
At the next level, the lower predicate calculus,
propositions can also be built from other elements, such as 
numbers or group elements $m,n$.
One then needs a predicate such as ``=''
that makes a proposition $m=n$ out of two numbers or group elements.
The variables $m,n$ themselves are defined through their own axioms
such as the Peano axioms in the case of numbers;
in addition there are axioms that define how to
make a new number out of two numbers (e.g. via $+,\cdot$)
or a new group element out of two group elements (via $*$). 

\vskip3.7cm

 {}\epsffile[-5 5 0 0]{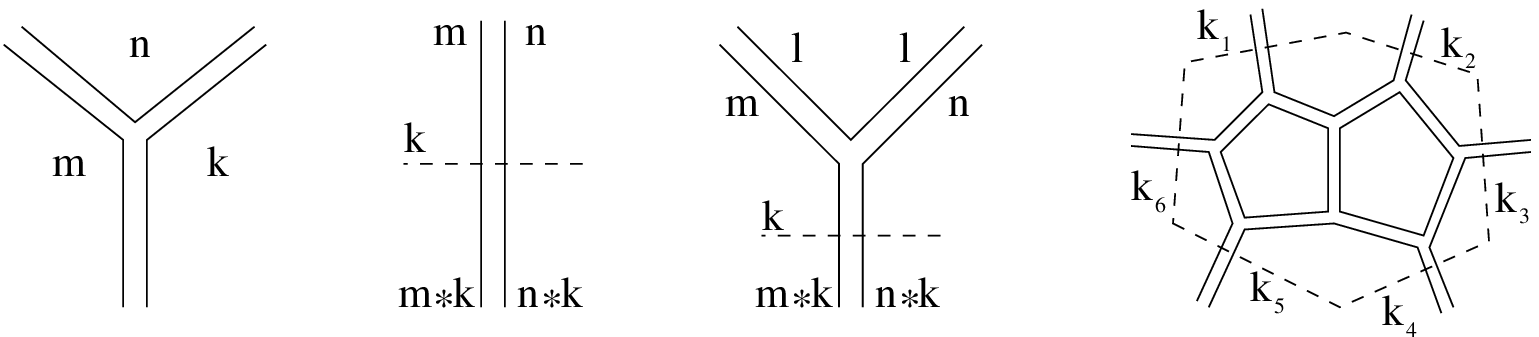}

\vskip9mm
\begin{center} {\small Figure 7: Axioms and proofs with predicates.}
\end{center}

The open problem to be discussed
 is how to represent all these new axioms in terms of
Feynman diagrams of some field theory. Here I will only comment on 
axioms involving the predicate.
The latter is generally required to 
obey the transitivity axiom
$$(m=n\ \wedge\   n=k )\ \supset\ \  m=k .$$
This axiom is graphically represented in figure 7 (left).
It is precisely the axiom that has been discussed
in the toy model of section 2.
So the proofs
that can be constructed with the transitivity axiom alone are
again the Feynman diagrams of a zero--dimensional SO($N$) theory.
The dual graphs are
triangulated string world--sheets with boundary,
 as for the propositional calculus.

\vskip2mm
\noindent 
$\underline{\hbox{Strings on group spaces}}$
\vskip1mm

\noindent
Next, I would like to argue that, after including the other axioms, the 
space of values of the variables $m,n,...$
becomes an embedding space for the boundary of the
 string world--sheet. To this end, let
us assume that $m,n,k,...$ are elements of a $d$--dimensional
Lie group $G$ and denote group 
multiplication by ``$*$'': $m*k\in G$. Let us graphically
represent 
 one of the
standard axioms,
$$(m=n)\ \ \ \supset\ \ \ (m*k\ =\ n*k)\ ,$$
as shown in figure 7 (second graph), where the dashed line
denotes multiplication with $k$.
The third graph in figure 7 then represents a proof of 
$(m=l)\wedge(l=n)\supset (m*k=n*k)$.
Figure 7 (right)
shows a more general proof in which group elements $k_i$ are associated 
with each boundary edge of the
dual triangulation. This defines a map from
the boundary to the $d$--dimensional 
group space. Likewise, the bulk of the triangulated string world--sheet
can be mapped onto group space.
 One may hope that a suitable
continuum limit exists, where the boundary
becomes a continuous curve that is embedded
in the group space. The boundary 
would then correspond to 
a correlation function of the corresponding 
 logical quantum field theory
that resembles a Wilson loop.
Like the loop, this theory
 would 
live in $d$ dimensions.

\vskip2mm
\noindent 
$\underline{\hbox{D--branes}}$
\vskip1mm

\noindent
If the boundary lives in $d$ embedding dimensions,
then the bulk of the world--sheet actually lives in $d+1$ dimensions.
This is a well--known feature of string representations
of large-$N$ field theories (such as our logical
quantum field theories): there is a new embedding dimension
that arises from the world--sheet conformal factor \cite{pol}. 
The $d$--dimensional hypersurface on which the world--sheet
ends is a so--called ``D--brane''
of the $d+1$--dimensional string theory.
The bulk of the world--sheet represents proofs,
and the boundary represents theorems; so the ensemble of {\it both} proofs and 
theorems would be represented by
 the ensemble of closed strings that are allowed to end on
 a D--brane of codimension 1.

Of course, once $d>1$, one really needs superstrings in order to avoid
the tachyon problem of the bosonic string. Do superstrings arise in logic?

\vskip2mm
\noindent 
$\underline{\hbox{Superstrings?}}$
\vskip1mm

\noindent
In the above, proofs built from the transitivity axiom 
and  proofs built from axiom 4 (\ref{mick}) have both been represented by
double line graphs that are dual to triangulated surfaces.
The main difference is that the individual lines represent
numbers or group elements $g\in G$ in one case, and
 propositions $p\in\{true, false\}$ in the other case.
Next we should combine both types of proofs. This requires including
the rule of substitution and must therefore be left for the future.
One hope is that numbers (or
group elements)
and propositions become superpartners,
so that the logical quantum field theory lives in a superspace 
-- or equivalently,  
 is a supersymmetric large--N theory that lives on a D--brane
of some dual superstring theory.

{}\subsubsection*{4.3. Nonperturbative Field Theory and Undecidable Theorems}

Let me finally mention
 another original motivation for defining
``logical quantum field theories''. At least before
adding the rule of substitution, they provide a map
from theorems $T$ into the space of functions $f_T(\beta,N)$ of a real parameter $\beta$
and an integer parameter $N$ 
 -- the corresponding correlation functions. These functions are 
\begin{enumerate}
\item
zero, if the theorem is false.
\item
$f_T(\beta,N)\ =\ \sum_{p(T)} e^{-\beta L(p)}f_p(N)$ if the theorem 
is true and can be proven,
\end{enumerate}
where $L$ is the length of a proof $p$ of a theorem $T$, and $f_p$ is
a power of $N$.
What about theorems that are true but cannot be proven within
a given formal system?
Such theorems are well--known to exist in number
theory, and can already exist in 
truncations of the propositional calculus.
If there is no proof of a theorem, 
this means that there is no Feynman diagram
that contributes to the corresponding correlation function.
So the correlation function
 must be zero in a perturbation expansion in $\beta$.
But this leaves open the possibility that the correlation function is
\begin{itemize}
\item[3.]
nonperturbatively nonzero, if the theorem is true but cannot be
proven.
\end{itemize}
In other words, the logical quantum field theory
 might know more about theorems 
than the formal system: 
it might know about theorems that are undecidable within this formal system.
This is a version of the standard fact that
a field theory knows more about correlation functions
than its Feynman diagrams: it knows about nonperturbative effects
(although nonperturbative effects do show up in the Feynman diagram
expansion
in terms of singularities of the Borel transform).

If such a relation between nonperturbative
field theory and undecidable theorems exists, it would raise
a fascinating question: are there
dualities in logic that interchange provable and unprovable theorems
and that correspond to the strong--weak coupling dualities known from 
string theory? 
Indeed, theorems that cannot be proven in one
formalization of a given branch of mathematics
 may be provable in another formalization, and vice versa.
These two formalizations
would then correspond to different perturbation expansions
of the same logical quantum field theory.
In switching between them,
the functions $f_T(\beta,N)$  that we have associated with theorems
would undergo a duality transformation.
This is planned to be investigated in the future.

{}\subsection*{5. Possible interpretation}\scs{5}

Let me conclude by first stating from a slightly
different perspective what we have done. We have considered
the sets of proofs in formal
systems as statistical--mechanical ensembles, and we have discussed
their statistical--mechanical properties. That is, we have considered
partition functions of the type
\ba{Z(\beta,\ ...)}\ \sim\ \sum_{p} e^{-\beta L(p)\ +\ ...}\ 
,\la{ensemble}\ea
where $L$ denotes the length of a proof $p$, measured in axiomatic steps,
and the dots represent possible other parameters in the action (in
addition to $\beta$).
One could now interpret $\beta^{-1}$ as a 
temperature, and define thermodynamic quantities for the ensemble of proofs,
such as the specific heat 
$$ c(\beta)\ \sim\ \partial_\beta^2\ \log Z(\beta)\ .$$
One could then study 
the specific heat as a function of the temperature $\beta^{-1}$,
and ask questions such as whether there
is a second order phase transition at some critical temperature.

At least for simple subsystems of the standard propositional
and predicate calculus, we have mapped
the proofs onto triangulated random
surfaces. These surfaces are made of logical axioms in the sense that each
triangle represents one of the axioms from which the proof is constructed.
If the temperature $\beta^{-1}$ is fine--tuned ``by hand'' to a critical
value, then indeed a second--order phase transition can be studied where the 
triangulated surfaces can be interpreted as continuum string world--sheets.
(\ref{ensemble}) then has an interpretation
as a string partition funtion.
For more general cases, it has been speculated (in subsection 4.2) that
 the ensemble of theorems and their proofs
becomes the ensemble of string world--sheets whose boundaries
live on a D--brane of codimension 1.

The question then arises whether these ``logical string world--sheets''
can be identified with the string world--sheets that the real world is
assumed to be made of (after switching to Minkowskian signature). 
This question will be discussed next.

\vskip2mm
\noindent
$\underline{\hbox{The ``Mathscape''}}$
\vskip1mm

\noindent
Let me first apologize for the philosophical and therefore vague character 
of the following remarks. Even if they are vague,
I want to mention them as a main motivation for trying to make
 a connection between string theory
and logic.

Identifying the logical with the physical strings seems to require that we
take a ``platonic'' attitude towards mathematics, regarding it
as an abstract reality that exists independently
of mathematicians, in the sense that mathematicians can discover it
by means of logical reasoning but they cannot change it. 
E.g., we
 certainly cannot change the fact that 335149 is a prime number,
but we can find out that it is true. 

The theorem ``335149 is a prime number'' is only one out of an
infinite number of facts that can be discovered: 
given the axioms of number theory,
what are the theorems 
that can be proven with them? Moreover,
how long is the shortest proof of a
given theorem? How many distinct proofs are there with given length $L$?
More generally, what are the
properties of the ensemble of theorems and their proofs?
Let us call this ensemble of theorems and proofs  in general formal
systems the
``mathscape''.\footnote{Modifying R. Rucker's
term ``mindscape'' \cite{ruc}.}
Let us take the viewpoint that it makes no
 sense to further ask where the mathscape comes from -- 
like the list of prime numbers, it is ``simply there'' --
but that it makes sense to study its properties.

Regarded as a statistical--mechanical system,
what does this mathscape ``look like''? To describe it, it first
of all seems sufficient to consider the 
set of theorems and
proofs in number theory: following \cite{tur}, number theory
is ``sufficiently powerful'' in the sense that theorems and proofs
in all other formal systems can be mapped onto theorems and proofs
in number theory.

We have considered the beginnings of number theory, namely
the propositional calculus and simple extensions of it. In these cases
we have argued that the mathscape resembles
 the real world: it contains
strings and (perhaps) membranes. Moreover, these strings are 
automatically first--quantized:
first quantization of the logical strings simply reflects
the standard relation between field theory and statistical mechanics,
applied to the ensemble of proofs.
If the string representation of general proofs in number theory 
could be derived rigorously, this would support
 the hypothesis that
the ``real world'' {\it is} the
mathscape, or at least  some basic part of it.

\vskip2mm
\noindent
{}$\underline{\hbox{Two objections}}$
\vskip1mm

\noindent
The idea that the physical strings are logical strings  of the
type discussed here 
 may  raise eyebrows for many reasons. 
Let me try to reply to two of the immediate objections one might have:
\begin{enumerate}
\item
These logical strings are ``unphysical'' -- they are
``abstract'' strings that live in an abstract space of logical proofs. 
The reply is that
we -- the observers -- would ourselves be unphysical 
in the same sense, since we are also made
of strings and therefore axioms, and therefore we would live in exactly
 the same
abstract space. And for abstract observers, 
abstract strings may be as ``real'' 
as physical strings are for physical observers.
Similar remarks hold for the
computer--generated observers in computer--generated universes
discussed in \cite{jur}, and are also made in \cite{teg}.
\item
 {\it Who} fine--tunes $\beta$ in partition functions of the type
(\ref{ensemble}) to the critical value?
 If $\beta$ were not fine--tuned, we would
not see continuum strings  in the real world, but, if anything,
discretized strings.
The reply to this is that this fine--tuning problem is nothing but
the tachyon problem of bosonic string theory: at least
in the simple models discussed here, $\beta$ can be regarded as
the world--sheet cosmological constant, i.e., as a tachyon zero mode. So
it remains to get rid of the tachyon, e.g. by supersymmetry
or in whatever way the hypothetical QCD string gets rid of it.
\end{enumerate}

\vskip2mm
\noindent
$\underline{\hbox{Renormalization group flows}}$
\vskip1mm

\noindent
Once $\beta$ in (\ref{ensemble}) is
near its critical value, 
the statistical--mechanical properties of the ``mathscape''
 can be studied using
renormalization group methods. The simple formal systems represented in
sections 2 and 3 
 contain only two parameters, $N$ and $\beta$.
In the case of more general predicate
calculi, where one has higher--dimensional string embedding spaces (see
subsection 4.2), the action will contain many parameters,
corresponding to all the string fields.
These paramenters
will flow as one considers the mathscape at larger and larger
scales in the sense of larger and larger proofs. Renormalization
group trajectories correspond
to $\phi$--dependent solutions of string theory, where $\phi$
denotes the additional embedding dimension that arises from the conformal factor
of the world--sheet \cite{wadia}. 

As in \cite{st} (which was already motivated in part
by these ideas, but where $\phi$ was ``time''), one would flow from
some UV fixed point to some stable IR fixed point.
The IR fixed point of the mathscape would
correspond to a stable string vacuum.
 What is the UV fixed point?
It would describe the ``bare'' mathscape.
It would be interesting to find solutions where the UV fixed
point corresponds to a topological string theory.

\vskip2mm
\noindent
$\underline{\hbox{Relation to previous papers}}$
\vskip1mm

\noindent
To conclude, let me 
relate the remarks in this section
to two previous papers \cite{jur,teg}.

In \cite{jur}, J\"urgen Schmidhuber studies the ensemble of bit sequences 
that can be computed by computer programs. This 
includes the ensembles of proofs and theorems in formal systems,
since those can be encoded as bit sequences that can be computed.
Our universe is interpreted in terms of such bit sequences.
 One of the aspects of \cite{jur} that has no analog here
is that bit sequences are weighted by their
Kolmogorov complexity, rather than by the length of the computation
(whose analog, the length of proofs, has led us to string theory here).

Tegmark \cite{teg} considers 
a step--wise generalization of properties 
of the observable world, from varying the parameters of the Standard Model
to making space--time discrete.
It is concluded that the most general universe should 
be related to the most general logical calculus 
in a sense that does not seem to be specified.
The present note specifies such a relation between logic
and particle physics by suggesting an explicit map from proofs
in formal systems to first--quantized string world--sheets.

{}\subsubsection*{Acknowledgements}

\noindent
I would like to thank P. Mayr and J. Schmidhuber for encouragement,
A. Beliakova, M. Hutter and B. Scarpellini for comments on the manuscript,
and the audience at CERN for a lively seminar.


\begin{thebibliography}{99}
\bibitem{pla}
This table of the four elements is due to the Greek philosopher
Empedocles (5th century B.C.). The interpretation in terms of
four of the five regular polyhedra
appears in Plato's dialogue ``Timaeus''; the role of the fifth polyhedron
seems less clear.
\end{thebibliography}
\end{document}